# Possible realization of three-dimensional quantum spin liquid behavior in HoVO$_4$

Dheeraj Ranaut[1], Shivprasad S. Shastri[2], Sudhir K. Pandey[2] and K. Mukherjee[1,*]

[1]School of Basic Sciences, Indian Institute of Technology Mandi, Mandi 175005, Himachal Pradesh, India

[2]School of Engineering, Indian Institute of Technology Mandi, Mandi 175005, Himachal Pradesh, India

*Email: kaustav@iitmandi.ac.in

**Abstract**

The study of geometrically frustrated magnetic systems with unusual crystal field ground states offers a possibility of realizing the new aspects of physics of disordered systems. In this study, we report our results of structural, magnetic susceptibility, heat capacity measurements, along with density functional theory calculations on HoVO$_4$; a compound in which the presence of a distorted kind of HoO$_8$ polyhedral leads to multiple magnetic interaction paths. The observed broad maximum below 10 K in the temperature response of DC susceptibility curves implies the presence of short-range correlations. AC susceptibility rules out the possibility of any kind of spin freezing. Temperature dependent heat capacity measurement at zero field indicate towards the absence of long-range ordering, along with the presence of a broad maximum centered around 14 K. The residual heat capacity exhibits a characteristic power-law ($T^\alpha$) behavior with the exponent $\alpha$ nearly equal to 2, which is analogous to that observed for other 3D quantum spin liquid systems. The density functional theory calculations signify the presence of dominant second and third nearest neighbor interactions, which in turn lead to magnetic frustration in our system. Our investigations suggest that HoVO$_4$ can be a candidate for realizing a 3D quantum spin liquid state.

# 1. Introduction

The presence of geometrical magnetic frustration in a system leads to a strong competition between spin-spin interactions and results in a plethora of exotic ground states; such as spin glass, spin ice and spin liquid (SL), which are under extensive investigations in recent times [1]. In these systems, incompatibility of the spins to satisfy the nearest-neighbor interactions, results in absence of any kind of conventional long-range ordering (LRO) [2-4]. In the past two decades, several new magnetic materials have been proposed to exhibit SL states (see recent reviews, Refs. 3-5). Low spin values and quasi low dimensionality are considered to be the main ingredients for the observation of SL state in a system. Most of the proposed quantum spin liquids (QSL) are materials with spin 1/2 moments lying on the quasi-low dimensional structures. These structures include two dimensional (2D) triangular and kagome lattice. Examples of triangular lattice include some organic compounds κ-(BEDTTTF)$_2$Cu$_2$(CN)$_3$ [6] and EtMe$_3$Sb[Pd(dmit)$_2$]$_2$ [7], along with some rare earth oxides YbMgGaO$_4$ [8] and TbInO$_3$ [9, 10]. While Kagome materials include inorganic compounds such as Herbertsmithite (ZnCu$_3$(OH)$_6$Cl$_2$) [11] and BaCu$_3$V$_2$O$_8$(OH)$_2$ [12], where $S =1/2$ moments form a highly frustrated network of corner-sharing triangles. The QSL states in these types of materials may follow either fermionic or bosonic statistics having exotic elementary excitations and can have gapped or gapless energy dispersion behavior [5]. However, the field of SL is still an emerging topic in three-dimensional (3D) spin systems.

Among the condensed matter physicist, it is of great interest to investigate whether a QSL state can be achieved in a 3D spin system. This issue is very important as quantum effects are significantly reduced on approaching a 3D system from low dimensional systems. For a 3D spin system to exhibit spin liquid like behavior, strong magnetic frustration is an important ingredient, as it enhances the quantum fluctuations. SL state has been explored very well in pyrochlore lattice, which is a well-known example of frustrated 3D network of corner-sharing tetrahedral [13-15]. Further, 3D networks of corner-sharing triangles exemplified by the hyperkagome lattice are also expected to support the SL behavior in the presence of isotropic antiferromagnetic (AFM) interactions [16, 17]. Three representative systems in the family of corner-shared triangles are Gd$_3$Ga$_5$O$_{12}$ [18, 19], Na$_4$Ir$_3$O$_8$ [20] and PbCuTe$_2$O$_6$ [21, 22]. The latter among these show a field induced QSL behavior due to the dominance of second near-neighbor (NN) interaction which forms a frustrated hyper-kagome network. As far as the 4$f$ rare earth oxides are concerned, the QSL state has been reported in TbInO$_3$ [9, 10] and YbMgGaO$_4$ [8], having 2D triangular lattice structure. For 3D network, Gd$_3$Ga$_5$O$_{12}$ [18, 19] and Ce$_2$Zr$_2$O$_7$ [23] are among the QSL candidates which belong to rare earth family. The compound TbInO$_3$ seems to be an interesting candidate because of the unusual Schottky anomaly present in heat capacity at a temperature relatively higher than that for other QSL systems. This observation can be described by adopting a slightly different crystal field (CF) scheme, rather than a conventional two-level CF scheme. However, the search for a potential 3D rare earth (4$f$) QSL compound is still in progress.

Herein, we reintroduce a rare earth compound HoVO$_4$, a Van Vleck paramagnet which belongs to the family of rare earth orthovanadates, having chemical formula RVO$_4$ (R = rare-earth). The compounds of this series show interesting magnetic and optical properties due to the presence of indirect super-exchange magnetic interaction and 4$f$ electron-phonon coupling [24-28]. Recently, this series is in focus, as a few compounds show exotic properties. In TmVO$_4$, magnetic field tuned ferroquadrupolar quantum phase transition is reported [29, 30] and anisotropic nematic fluctuations are found to exist above this transition [31]. Two other members of the family reports the presence of field tuned quantum criticality in DyVO$_4$ [32] and signatures of effective spin – ½ ground state in CeVO$_4$ [33]. Spectroscopic and crystal field excitation studies on HoVO$_4$ reveal the presence of a non-magnetic singlet ground state, along with first excited doublet at 30 K. This first excited doublet shows a zero-field splitting of ~ 3.3 K at low temperatures. Further, the second excited singlet and third excited doublet are found to be accidently degenerated around 60 K [34, 36, and 38]. This CF schemes in HoVO$_4$ is expected to strongly affect the specific heat as almost similar kind of CF scheme is analyzed in TbInO$_3$ [9, 10]. One more point of similarity is the presence of non-Kramers ion, Tb$^{3+}$ (4$f^8$) in TbInO$_3$ and Ho$^{3+}$ (4$f^{10}$) in HoVO$_4$. This result in the complete lifting of the ground state degeneracy by the CF, leading to a non-magnetic singlet ground state. These observations motivated us to reinvestigate HoVO$_4$ in terms of SL behavior.

Hence, in this work, we have adopted temperature and magnetic field dependent heat capacity study, along with basic magnetic measurements and computational studies (to calculate the magnetic interactions) as major tools to reexamine HoVO$_4$. This compound crystallizes in tetragonal structure and forms a 3D network of Ho atoms. The presence of distorted kind of HoO$_8$ polyhedron results in different magnetic interactions, depending on the Ho-Ho distance and the super-exchange paths. The temperature dependent magnetic susceptibility reveals the presence of antiferromagnetic correlations, along with absence of LRO down to 1.8 K. It also exhibits a broad hump centered around 5 K, implying the onset of short-range correlations. The zero field heat capacity data exhibits a broad hump around 14 K, along with absence of any LRO down to 1.8 K. The value of the exponent (~ 2.2) obtained from the power-law ($\sim T^{\alpha}$) dependence of residual heat capacity lies in the range of other 3D QSL [20, 21]. Magnetic interaction parameters obtained from the density functional theory (DFT) calculations signify the presence of strong magnetic frustration resulting in a possible QSL state.

## 2. Methods

Polycrystalline sample of HoVO$_4$ has been prepared by the conventional solid state reaction method by using high purity Ho$_2$O$_3$ and VO$_2$ from Sigma Aldrich by using the method described in [32]. Room temperature powder X-ray diffraction (XRD) is performed in the range (10°- 90°) using Rigaku diffraction with Cu K$\alpha$ ($\lambda$ = 1.54) radiation. The crystal structure is refined by the Rietveld method using FullProf Suite software. Magnetic field ($H$) and $T$ dependent magnetic measurements in the temperature range 1.8-300 K are performed using the Magnetic Property Measurement System (MPMS) from Quantum Design, USA. Physical Property Measurement System (PPMS), Quantum Design, USA is used to measure $T$ dependent heat capacity ($C$) at different fields up to 70 kOe.

Exchange parameters ($J_{ij}$) are calculated for the spin Hamiltonian:

$$H = -\sum_{i>j} J_{ij} \boldsymbol{S}_i \boldsymbol{S}_j \quad \ldots\ldots\ldots (1)$$

where $S_i$ and $S_j$ corresponds to the spin of Ho atoms at $i$ and $j$ sites, respectively. The spin-polarized total energy calculations for different defined magnetic structures are carried out using full-potential linearized augmented plane wave (FP-LAPW) method-based code Elk [38, 39]. The nature and magnitude of first three NN exchange parameters are calculated by mapping the total energy corresponding to different defined magnetic structures on the Hamiltonian described in equation (1). The crystal structure information for the calculations is used from our Rietveld analysis of experimental XRD data shown in Fig. 1. For the exchange-correlation, LSDA of Perdew-Wang/Ceperley-Alder is used [41]. The muffin-tin sphere radii values of 2.2, 1.65 and 1.43 Bohr are used for Ho, V and O atoms, respectively. A convergence criterion of 10$^{-5}$ Ha/cell is used for total energy in the self-consistent cycles. The total energy and moments of different magnetic structures are obtained with and without spin orbit coupling (SOC) contribution. For this purpose, original unit cell is extended twice in z-direction. Mainly, a k-mesh of size 6x6x3 is used in these calculations. The corresponding microscopic magnetic model is discussed further in more detail in the section 3.4.

## 3. Results

### 3.1 X-ray diffraction and structural features

Fig. 1 displays the obtained XRD, and the refined patterns of HoVO$_4$ with *2θ* ranging from 10° to 90° at room temperature. All the diffraction peaks can be indexed into tetragonal crystal structure having space group I4$_1$/amd (No. 141). The obtained lattice parameters from the Rietveld analysis are a = b = (7.122 ± 0.004) Å and c = (6.289 ± 0.007) Å with goodness of fit (GOF) ~ 2.14. The XRD pattern and obtained values of the lattice parameter match very well with previous literature reports on HoVO$_4$ [34, 35]. It crystallizes in tetragonal crystal structure, belonging to Zircon class of minerals. Its structure consists of isolated VO$_4$ tetrahedron (orange color) which share corners and edges with HoO$_8$ bisdisphenoids (grey color), as shown in Fig. 2 (a). Each Ho atom is bonded to four O atoms at (Ho-O$_c$ = 2.31(5)) Å and to other four O atoms at (Ho-O$_e$ = 2.46 (5)) Å. Here, the subscript c and e stand for compressed and elongated tetrahedra, respectively. So, the HoO$_8$ polyhedron is comprised of two interpenetrating HoO$_4$ tetrahedra and can be treated as a distorted kind of geometry. One of the tetrahedra is elongated having larger Ho-O$_e$ distance and the other is compressed, having smaller Ho-O$_c$ distance. The magnetic interactions take place through super-exchange path (Ho – O – Ho) and super-super exchange path (Ho – O – V – O – Ho). Due to the presence of two HoO$_4$ tetrahedra, different exchange paths are expected based on the nearest Ho-Ho distances. Fig. 2 (b-e) depicts the possible networks/geometry of spins corresponding to different NN interactions. The first NN distance between Ho atoms is about (3.893 ± 0.002) Å and exchange coupling takes place through the path Ho – O – Ho. These first NNs form a 3D network of spins (shown in Fig. 2 (b)) and the interaction is termed as $J_1$. The second NNs interaction ($J_2$) correspond to Ho-Ho distance of (5.909 ± 0.007) Å. This interaction is mediated through the path (Ho – O$_e$ – V – O$_c$ – Ho). Fig. 2 (c) shows that these atoms form a 3D network of corner-sharing tetrahedra with one Ho at the center of each tetrahedra. $J_3$ is related to the third NN atoms with a Ho-Ho distance of (5.937± 0.0012) Å. It is intervened through the super-exchange path given by: (Ho – O$_c$ – V – O$_c$ – Ho). These third NN Ho atoms form an edge-sharing network of polyhedron/cuboid with one Ho atom positioned at center (Fig. 2 (d)). $J_2$ and $J_3$ are expected to jointly induce magnetic frustration in this system, as they form frustrated network of spins. The fourth NNs distance between Ho atoms is about (6.288 ± 0.011) Å and the exchange interaction ($J_4$) between them takes place through the path (Ho – O$_e$ – V – O$_e$ – Ho). These Ho atoms form linear chains in which the interaction takes place through the above-mentioned path (shown in the Fig. 2 (e)). These chains are interlinked via $J_1$ path. One more important thing to note is that these Ho atoms forming chains are separated from each other via non-magnetic tetrahedra VO$_4$ which is an important aspect to achieve low-dimensional magnetism. As the bond length is large, the chain coupling ($J_4$) is expected to be weaker than the other magnetic interactions. In order to get an insight on the possible magnetic interactions in HoVO$_4$, we have performed temperature dependent DC and AC susceptibility measurements at various applied external magnetic fields. This is further followed by the computational study on the evaluation of these different magnetic interactions ($J_i$).

## 3.2 Magnetic Susceptibility

The temperature dependence of DC magnetic susceptibility ($\chi_{DC}$) under zero-field cooled (ZFC) protocol at 100 Oe is shown in the left panel of the Fig. 3 (a). With decrease in temperature, $\chi_{DC}$ increases continuously and the signature of long-range ordering is absent down to 1.8 K. Our observation matches well with previous literature reports where no magnetic ordering is reported for HoVO$_4$ [36, 37]. The right panel of Fig. 3 (a) displays the temperature dependence of inverse DC susceptibility ($\chi_{DC}^{-1}$). This curve is fitted using the equation: $\chi_{DC} = \chi_0 + C/(T - \theta_{CW})$, where $C$ is the Curie constant and $\theta_{CW}$ is CW temperature. $\chi_0$ contains the contribution from the core diamagnetic susceptibility and temperature independent Van Vleck paramagnetic susceptibility. It can be noted from the figure that above 25 K, $\chi_{DC}^{-1}$ curve is fitted well with the CW law (shown by the red solid line). Below this temperature, a non-linear deviation is observed which may be due to CF effect. The obtained values of $\chi_0$, $C$ and $\theta_{CW}$ are ~ (1.15 × 10$^{-3}$ ± 0.07 × 10$^{-3}$) emu/mole-Oe, ~ (15.37 ± 0.02) emu/mol-Oe-K and ~ (-12.36 ± 0.09) K, respectively. The experimentally calculated effective magnetic moments $\mu_{eff}$ (= $2.83\sqrt{C}$ $\mu_B$ ~ 11.09 $\mu_B$), is slightly larger than the theoretically calculated value of $\mu_{eff}$ = 10.61 $\mu_B$ for Ho$^{3+}$. This kind of large experimental effective moment has also been reported in other frustrated magnetic systems [42]. The negative value of $\theta_{CW}$ indicates the dominance of AFM interactions in this compound. Generally, for an ideal AFM system, one would expect -$\theta_{CW}$ ~ $T_N$ (AFM transition temperature). But in our system, the presence of large $\theta_{CW}$, along with absence of any sharp transition feature down to 1.8 K, signifies the presence of magnetic frustration. The inset of Fig. 3 (a) represents the $\chi_{DC}$ curve in low temperature region (below 25 K) under both ZFC and field-cooled (FC) protocols at 100 Oe. Both curves superimpose on each other. Below 10 K, $\chi_{DC}$ exhibits a broad maximum centered around 5 K followed by an upturn at 2.2 K. It is to be noted here that this unusual behavior of $\chi_{DC}$ (observed below 10 K) has not been studied in any previous studies. Generally, a broad peak in bulk susceptibility well below $|\Theta_{CW}|$ arises due to short range correlations and this feature has also been reported in many frustrated spin systems [43, 44]. In order to investigate the magnetic correlations below 10 K, $T$ dependent AC susceptibility at different frequencies 31-831 Hz with AC amplitude 2 Oe is measured. Fig. 3 (b) depicts the real part of the AC susceptibility ($\chi'$) at different frequencies for zero applied DC field. The curves do not exhibit any frequency dependent peak within the above-mentioned frequency range, implying the absence of any kind of spin freezing. Hence it can be said that the short-range correlations observed below 10 K do not correspond to any kind of spin freezing or spin blocking. So, the presence of broad hump corresponding to the short-range correlations and large value of $\Theta_{CW}$ provide sufficient experimental signatures of magnetic frustration in our system $HoVO_4$. Along with this, the inset of Fig. 3 (b) shows the $T$ dependent $\chi'$ at different applied DC fields at 331 Hz. It can be clearly seen that with increasing field, the $\chi'$ decreases, and approaches a constant value as $T$ is lowered to 1.8 K. This kind of behavior is not unusual as has also been reported in QSL Kitaev magnet $Cu_2IrO_3$, where it has been attributed to the emergence of low-energy excitation [46]. Along with this, in $Cu_2IrO_3$, the DC susceptibility approaches a constant value as $T$ is lowered [47]. Furthermore, Fig. 3 (c) shows the $T$ response of $\chi_{DC}$ at different $H$ in log scale. For $H \geq 10$ kOe, the green arrow shows that $\chi_{DC}$ approaches a finite constant value as $T \rightarrow 0$. It acquires a finite value at 2.8 K for 10 kOe curve and this temperature corresponding to the initial point of constant value increases for higher fields. This temperature corresponds to 4.3, 6.4 and 8.8 K for 30, 50 and 70 kOe curves, respectively. This behavior is in accordance with DC field superimposed $\chi'$, which also attains temperature independent behaviour at low temperatures. Similar kind of behavior had also been reported in another QSL candidate as well [45]. The next section focuses on the $T$ dependent $C$ at different applied $H$, as it will provide further insight about the ground state of this system.

### 3.3 Heat Capacity

Heat capacity acts as a sensitive tool to probe the low-energy excitations between the ground state and excited states of a system. $T$ dependent $C$ measurements are performed down to 1.8 K in different $H$ up to 70 kOe. To extract the residual heat capacity contribution $C_{res}$, the lattice part is estimated by fitting the raw data with an equation consisting of linear combination of Debye and Einstein terms [48]:

$$C(T) = mC_{Debye}(T) + (1-m)C_{Einstein}(T) \ldots\ldots\ldots (2)$$

where *m* describes the weight of the Debye term. $C_{Debye}$ and $C_{Einstein}$ are the Debye and Einstein contributions to the lattice heat capacity, respectively, which can be described by the equations given as:

$$C_{v\,Debye}(T) = 9nR\left(\frac{T}{\theta_D}\right)^3 \int_0^{\frac{\theta_D}{T}} \frac{x^4 e^x}{(e^x-1)^2} \quad \ldots\ldots (2\,(a))$$

$$C_{v\,Einstein}(T) = 3nR\left(\frac{\theta_E}{T}\right)^2 \frac{e^{\frac{\theta_E}{T}}}{(e^{\frac{\theta_E}{T}}-1)^2} \ldots\ldots (2\,(b))$$

where n is the number of atoms per formula unit, R is the universal gas constant, $\theta_D$ and $\theta_E$ are Debye temperature and Einstein temperature, respectively. A good fitting of the equation (2) is observed with the Debye temperature ~ (739 ± 18) K and the Einstein temperature ~ (146 ± 05) K, and *m* ~ (0.83 ± 0.01) for Debye term (shown in the red solid line in Fig. 4 (a)). From fitting, it can be concluded that lattice contribution is negligible below 40 K. The absence of any sharp peak down to 1.8 K rules out the possibility of LRO and signifies that the upturn seen in $\chi_{DC}$ around 2.2 K cannot be attributed to any kind of ordering in our system. From the obtained *C* data, different features are clearly visible corresponding to different temperature regimes. Above 40 K, the fitted curve corresponding to equation (2) (shown by the red solid line in Fig. 4 (a)) clearly signifies the dominance of lattice contributions in the defined region. For an intermediate temperature range (2.5 K – 30 K), *C* shows a broad hump centered at 14 K. This region can be more clearly visualized in Fig. 4 (b) which describes the residual heat capacity, $C_{res}$ ($C_{res} = C - C_{lattice}$) on a log-log scale. In an earlier report, the existence of this anomaly in *C* in $HoVO_4$ was ascribed to some local distortions in the lattice [35]. However, similar kind of broad hump in $C_{res}$ has also been reported in the systems having magnetically frustrated ground state which also includes Ho-based systems as well [49, 50, and 51]. Further, the inset of Fig. 4 (b) shows the plot of the derivative of $C_{res}$ (*T*) with respect to temperature. It shows a presence of an anomaly around 5 K, shown by the blue solid arrow which is further discussed later. For T ≤ 2.2 K, an upturn in $C_{res}$ is noted upon further cooling, as shown by the black arrow in the Fig. 4 (b), with $C_{res}$ vs *T* on a log-log scale. This feature is believed to originate from the nuclear Schottky anomaly of Ho ions [52]. Moreover, this trait can be more clearly visualized in the Fig. 4 (c) which represents *T* response of $C_{res}$ in the presence of different applied fields. From the yellow shaded region, it can be inferred that the increment in $C_{res}$ at low temperatures is dramatically increased in the presence of *H*. This indicates that the nuclear Schottky anomaly becomes significantly strong in *H* as the Ho electronic moments become polarized along the field direction and thus produce a static mean field to which the nuclear spins are exposed [52]. Earlier experimental reports on HoVO$_4$ suggested that the Ho electronic spin moments do not order down to mK range, but the Ho nuclear spin moments show AFM ordering at 4.5 mK which is related to the nuclear Schottky anomaly [37].

In addition, from the Fig. 4 (c), it is observed that the broad anomaly observed in $C$ exhibits $H$ dependence. For 10 kOe curve, the position of the broad maximum remains identical, and the curve shows almost same behavior throughout the full $T$ range. For $H \geq 30$ kOe, the hump slightly broadens and shifts toward higher temperature. This kind of behavior is typically observed in the systems having doublet states and its splitting in energy takes place due to Zeeman effect [9]. In our system, application of $H$ is believed to result in the splitting of first excited doublet and third excited doublet. This energy difference between different levels further increases with increasing $H$. Along with this, non-linearity in the broad maximum with respect to $H$ is noted which may arise due to the polycrystallinity of the sample or there can be CF level crossings at low fields. Furthermore, the inset of Fig. 4 (c) shows the entropy change ($\Delta S$), obtained by integrating $C_{res}/T$ vs $T$ curve at zero field. From here, it is noted that $\Delta S$ reaches a saturation value of 10.6 J/mole-K at 50 K. This has been further discussed in detail in the section 4. We now focus on the $T$ – region of our interest where the $C_{res}$ shows some anomalous behavior.

As mentioned earlier, the $T$ response of the derivative of $C_{res}$ at zero field shows an anomaly around 5 K. This temperature coincides with the $T$ at which $\chi_{DC}$ shows the evidence of short-range correlations; implying the anomaly in $C_{res}$ is of magnetic origin. This further signifies the presence of some additional contribution to $C$ below 5 K, which can neither be ascribed to the broad hump anomaly centered at 14 K nor to the nuclear Schottky anomaly at lower temperatures. To explore the nature of magnetic excitation in this region, we have fitted $C_{res}$ data with the power law ($C_{res} \sim T^\alpha$). The red dash line in the Fig. 4 (b) shows the power-law behavior with an exponent $\alpha \sim (2.21 \pm 0.13)$. The deviation from the experimental data at low temperature is due to the presence of nuclear Schottky anomaly. The obtained value of $\alpha$ agrees very well with the observed value for other reported 3D QSL systems like Na$_4$Ir$_3$O$_8$ [20, 45], cubic-Ba$_3$NiSb$_2$O$_9$ [53] and PbCuTe$_2$O$_6$ [21]. For Na$_4$Ir$_3$O$_8$, its value was found to be around 2.4, while for cubic-Ba$_3$NiSb$_2$O$_9$, the observed value of $\alpha$ was about 2. For PbCuTe$_2$O$_6$, it was found to be around 1.9 for $H \geq 80$ kOe. From here, it can be concluded that our system HoVO$_4$ shows the power-law dependence of $C_{res}$ and the obtained value of the exponent lies in the range to that observed for other 3D QSL systems. So, our magnetic measurements along with the unusual power-law behavior in $C_{res}$ indicate towards the possible QSL ground state in HoVO$_4$. For the existence of such state in our system, it is expected that the magnetic couplings $J_2$ and $J_3$ (which forms frustrated 3D networks) should dominate, as magnetic frustration is an important precursor for QSL behavior in 3D spin systems. Hence, in order to get an insight on these different magnetic interactions, we have proposed a magnetic model which is discussed in the next section.

### 3.4 Microscopic magnetic model

The QSL behavior proposed on the basis of our experimental observations sounds quite baffling, considering the 3D nature of the crystal structure of HoVO$_4$. So, in order to justify this behavior, we have evaluated the individual magnetic interaction constants using DFT calculations. It is to be noted here that First-principles calculations on HoVO$_4$ was reported by Moussa et al., [54]. But this study was more focused on examining the electronic and magnetic ground state of the system and do not report about evaluating the different magnetic interactions. The total energy from the DFT calculations includes the non-magnetic part ($E_0$) along with the magnetic part. So, in order to calculate exchange parameters $J_{ij}$, corresponding to first three NNs (i.e., $J_1$, $J_2$ and $J_3$), we need at least four different equations described by the Hamiltonian in equation (1). In our magnetic model, we have defined four different magnetic structures namely, FM, AFM1, AFM2 and AFM3, having unique spin configurations and yielding different energies. These different magnetic structures are illustrated in the Fig. 5. One important thing to note here is that the unit cell is extended along the z-axis in order to include the effect of fourth NN Ho atoms. By mapping the energies corresponding to different magnetic structures into the Hamiltonian (1), four different equations can be easily extracted and are given as:

$$E_{tot}^{FM} = E_0 - 28J_1 - 20J_2 - 24J_3 - 16J_4 \quad \ldots\ldots \quad (3)$$

$$E_{tot}^{AFM1} = E_0 + 28J_1 + 20J_2 - 24J_3 - 16J_4 \quad \ldots\ldots \quad (4)$$

$$E_{tot}^{AFM2} = E_0 + 4J_1 - 4J_2 + 24J_3 - 16J_4 \quad \ldots\ldots \quad (5)$$

$$E_{tot}^{AFM3} = E_0 - 4J_1 + 4J_2 + 24J_3 - 16J_4 \quad \ldots\ldots \quad (6)$$

where, $E_{tot}^{FM}$, $E_{tot}^{AFM1}$, $E_{tot}^{AFM2}$ and $E_{tot}^{AFM3}$ are the total energies corresponding to different magnetic structures calculated by applying DFT formalism.

By applying simple mathematical algebra, the value of the magnetic interactions ($J_1, J_2$ and $J_3$) can be easily calculated and the obtained equations are given as:

$$E_{tot}^{AFM1} - E_{tot}^{FM} = 48J_1 + 40J_2 \ldots\ldots\ldots (7)$$

$$E_{tot}^{AFM3} - E_{tot}^{AFM4} = 8J_1 - 8J_2 \ldots\ldots\ldots (8)$$

$$E_{tot}^{AFM1} - E_{tot}^{AFM3} = 24J_1 + 24J_2 - 48J_3 \ldots\ldots (9)$$

From the above three equations, it can be inferred that the values of $J_1, J_2$ and $J_3$ are explicitly independent of the fourth NN magnetic interaction ($J_4$) and the model is adequate enough to complement our experimental observations of $J_2$ and/or $J_3$ being the dominating interaction path.

Our DFT calculations without spin-orbit coupling (SOC) yield the AFM3 to be the magnetic ground state with lowest energy and the values of magnetic interaction constants have been calculated using the above equations. The values of $J_1, J_2$ and $J_3$ are found to be 4.306 meV, -2.179 meV and -27.036 meV, respectively. The positive value of $J_1$ implies the presence of ferromagnetic interactions among the first NN Ho atoms, while negative values of $J_2$ and $J_3$ signifies the presence of antiferromagnetic correlations among the second and third NN Ho atoms. The large value of $J_3$ as compared to the other two parameters signifies the presence of magnetic frustration in the system. The described model seems to be quite reliable in order to consider only the magnetic interactions as none of the defined magnetic structures show the band gap at DFT level. Along with this, it is well known that DFT formalism with LSDA/GGA exchange correlation functional usually fails to predict the correct insulating ground state for strongly correlated oxide materials where 3d and/or 4f orbitals are partially occupied. The total density of state (TDOS) obtained for different magnetic structures is shown in the appendix. The unusually large values of the interaction parameters arise because of the large difference between the energies corresponding to different magnetic structures which may possibly contain contribution from factors like orbital angular momentum and SOC; as all these terms have a major involvement in 4f based systems.

Hence, we have also performed DFT plus SOC calculations as these calculations are expected to yield more accurate results (as compared to only DFT calculations) in rare earth-based systems. The obtained value of spin magnetic moment per Ho atom is 3.86 $\mu_B$ which is in good agreement with S = 2. The total energies corresponding to different magnetic structures follow the trend as: FM > AFM1 > AFM2 > AFM3, implying that the spin configuration AFM3 corresponds to the magnetic ground state. This is in agreement with DFT calculations as well as our experimental observations of AFM correlations among the spins. The values of $J_1$, $J_2$ and $J_3$ are found to be -3.98 meV, -16.98 meV and -20.67 meV, respectively. The negative value of $J_1$ signifies the presence of AFM correlations among the first NN Ho spins which is contradictory to simple DFT estimations but qualitatively consistent with the experimental analysis. Along with this, the significantly large negative values of $J_2$ and $J_3$ as compared to $J_1$ implies the dominance of second and third NN interactions, leading to magnetic frustration in $HoVO_4$. The TDOS for DFT + SOC calculations are shown in the appendix and it is noted that in analogy to DFT calculations, none of the magnetic structure shows the existence of band gap. Thus, our calculations of individual magnetic interactions indicate towards the dominance of $J_2$ and $J_3$, which forms the frustrated geometry of spins. This strong magnetic frustration, in turn, possibly results in a 3D QSL behavior in $HoVO_4$ at low temperatures.

## 4. Discussion

Since $Ho^{3+}$ is a non-Kramers ion having $4f^{10}$ electronic configuration, the ground state degeneracy can be completely lifted, and such systems are expected to have non-magnetic singlet ground state [9]. This argument is supported by earlier experimental observations [34, 36]. As discussed above, $HoVO_4$ crystallizes in the tetragonal structure in which Ho ion occupy the state with point group symmetry $D_{2d}$ (42m). In the tetragonal crystalline electric field, for $Ho^{3+}$ ion (4f electrons), the ground state multiplet $^5I_8$ splits into 9 singlets and 4 doublets. The CF GS in $HoVO_4$ is a non-magnetic singlet, with first excited doublet around 30 K. Further, the second excited singlet and third excited doublet are accidently degenerated around 60 K [33, 35]. Generally, for a two-level CF scheme, the value of $\Delta S$ is given by Rln $(1 + g_1/g_0)$ [55]. Here, R is the universal gas constant, $g_0$ and $g_1$ are ground state and first excited state degeneracy, respectively. From this equation, it can be observed that the expected $\Delta S$ for a two-level system with same degeneracy between ground state and first excited state (i.e., $g_0 = g_1 = 1$) is 5.76 J/mole-K and for a two-level system with singlet ground state ($g_0 = 1$) and doublet excited state ($g_1 = 2$), the value of $\Delta S$ is expected to be 9.13 J/mole-K. Thus, the significantly large value of $\Delta S$ for $HoVO_4$ (i.e., 10.6 J/mole-K) in comparison to the two-level system signifies the substantial contribution of the higher excited states to the CF scheme. Moreover, the shifting of the broad hump in $C_{res}$ on application of magnetic fields implies the splitting of the doublet excited states as a consequence of the Zeeman effect.

As far as HoVO$_4$ is considered, our experimental and computational studies provide sufficient evidences to support the possibility of 3D QSL behavior. Generally, most of the 3D spin systems exhibit ordered behavior in the absence of any special frustrated geometry, while ordering is prevented in one dimensional (1D) or 2D spin systems due to the presence of strong quantum mechanical fluctuations. In some 3D spin systems, conventional ordering is prohibited due to the strong geometrical frustration. Based on the different magnetic interaction paths and the frustrated networks formed by the second and third NNs, it appears that in HoVO$_4$ a strong magnetic frustration can be achieved. Our $C_{res}$ data reveals the presence of an anomaly around 5 K; the same temperature at which $\chi_{DC}$ yields a broad maximum. This signifies that the origin of anomaly in $C_{res}$ is related to the presence of magnetic correlations and provides additional contributions to the heat capacity. Even though the spin interactions in HoVO$_4$ are quite different from other 3D QSL systems like PbCuTe$_2$O$_6$ [21, 22] and Na$_4$Ir$_3$O$_8$ [20], it clearly exhibits evidences of spin liquid behavior, i.e. : absence of LRO as visible from the $\chi_{DC}$ and $C$ data at zero field; the presence of broad hump in $\chi_{DC}$ centered around 5 K, implying the onset of short range correlations; absence of spin freezing as evident from AC susceptibility measurement; a characteristic power-law behavior followed by $C_{res}$ curve below 5K with an exponent close to 2, which matches very well with other 3D QSL systems. From these experimental evidences, it is expected that the magnetic couplings $J_2$ and $J_3$ which forms frustrated 3D networks should dominate. This is conclusively established from our computational studies which give clear evidence of the dominance of magnetic interactions corresponding to second and third NNs. Thus, from our analysis, it can be inferred that HoVO$_4$ can be a potential candidate in order to study quantum spin liquid behavior in rare earth-based systems.

## 5. Conclusion

To conclude, the structure of HoVO$_4$ consists of a 3D network with various possible magnetic interaction paths. The $\chi_{DC}$ data shows the presence of strong AFM correlations, along with the absence of LRO which is reconfirmed by the absence of any sharp peak in $C$ at zero field. Below 10 K, $\chi_{DC}$ shows a broad maximum signifying the onset of short-range correlations which do not correspond to any kind of spin freezing as confirmed by AC susceptibility. $C_{res}$ follows a nearly $T^2$ behavior at zero field which matches well with other 3D SL systems. All these experimental observations are further supported by the computational study which shows the dominance of magnetic interactions corresponding to second and third NNs forming 3D frustrated networks of spins. From all the observations, it is suggested that $HoVO_4$ can be a example of 3D QSL spin systems belonging to the rare earth oxides family. Further, microscopic probes like NMR and inelastic scattering at various energy scales might unveil more about the spin excitations of the ground state in $HoVO_4$.

**Appendix: Total Density of states (TDOS) corresponding to different magnetic structures under DFT and DFT+SOC formalism.**

The spin-polarized total energy calculations are performed for four defined magnetic structures (FM, AFM1, AFM2 and AFM3) under DFT without and with SOC formalism, respectively. This exercise is performed to calculate the magnetic interactions corresponding to first three near neighbor Ho atoms. The TDOS obtained for these different structures under DFT and DFT+SOC formalism is shown in the Figs. A1 and A2, respectively. The Fermi level corresponding to Fermi energy is shown by the black solid lines in TDOS of each magnetic structure. None of these four structures show the presence of band gap for both formalisms, implying that the model is adequate enough for calculating the magnetic interactions.

**Acknowledgements**

The authors acknowledge IIT Mandi for the experimental facilities and financial support.

**Data availability statement**

All data that support the findings of this study are included within the article.

**References:**

1) Ramirez A. P., Strongly Geometrically Frustrated Magnets, 1994 *Annu. Rev. Mater. Sci.* **24** 453.
2) Lacroix C, Mendels P, and Mila F, Introduction to Frustrated Magnetism (Springer, Berlin, 2010).
3) Balents L, 2010 *Nature (London)* **464** 199.
4) Gardner J S, Gingras M J P and Greedan J E 2010 *Rev. Mod. Phys.* **82** 53.
5) Ramirez A P, 2008 *Nat. Phys.* **4** 442.
6) Yamashita S, Nakazawa Y, Oguni M, Oshima Y, Nojiri H, Shimizu Y, Miyagawa K and Kanoda K 2008 *Nat. Phys.* **4** 459.
7) Yamashita M, Nakata N, Senshu Y, Nagata M, Yamamoto H M, Kato R, Shibauchi T and Matsuda Y 2010 *Science* **328** 1246.
8) Li Y, Liao H, Zhang Z, Li S, Jin F, Ling L, Zhang L, Zou Y, Pi L, Yang Z, Wang J, Wu Z, and Zhang Q 2015 *Sci. Rep.* **5** 16419.
9) Kim J, Wang X, Huang F T, Wang Y, Fang X, Luo X, Li Y, Wu M, Mori S, Kwok D, Mun E D, Zapf V S and Cheong S W 2019 *Phys. Rev.* X **9** 031005.
10) Clark L, Sala G, Maharaj D D, Stone M B, Knight K S, Telling M T F, Wang X, Xu X, Kim J, Li Y, Cheong S W, and Gaulin B D 2019 *Nat. Phy.* **15** 262.
11) Helton J S, Matan K, Shores M P, Nytko E A, Bartlett B M, Yoshida Y, Takano Y, Suslov A, Qiu Y, Chung J H, Nocera D G, and Lee Y S 2007 *Phys. Rev. Lett.* **98** 107204.
12) Okamoto Y, Yoshida H, and Hiroi Z 2009 *J. Phys. Soc. Jpn.* **78** 033701.
13) Canals B and Lacroix C 1998 *Phys. Rev. Lett.* **80** 2933.
14) Moessner R and Chalker J T 1998 *Phys. Rev. Lett.* **80** 2929.
15) Savary L and Balents L 2017 *Phys. Rev. Lett.* **118** 087203.
16) Hopkinson J M, Isakov S V, Kee H Y, and Kim Y B 2007 *Phys. Rev. Lett.* **99** 037201.
17) Zhou Y, Lee P A, Ng T K and Zhang F C 2008 *Phys. Rev. Lett.* **101** 197201.
18) Schiffer P, Ramirez A P, Huse D A, Gammel P L, Yaron U, Bishop D J, and Valentino A J 1995 *Phys. Rev. Lett.* **74** 2379.
19) Deen P P, Petrenko O A, Balakrishnan G, Rainford B D, Ritter C, Capogna L, Mutka H and Fennell T 2010 *Phys. Rev. B* **82** 174408.
20) Singh Y, Tokiwa Y, Dong J, and Gegenwart P 2013 *Phys. Rev. B* **88** 220413(R).
21) Koteswararao B, Kumar R, Khuntia P, Bhowal S, Panda S K, Rahman M R, Mahajan A V, Dasgupta I, Baenitz M, Hoon Kim K, and Chou F C 2014 *Phys. Rev. B* **90** 035141.

22) Chillal S, Iqbal Y, Jeschke H O, Rodriguez-Rivera J A, Bewley R, Manuel P, Dmitry, Steffens P, Thomale R, Nazmul Islam A. T. M., Reuther J and Lake B 2020 *Nature Communication* **11** 2348.

23) Anish Bhardwaj, Shu Zhang, Han Yan, Roderich Moessner, Andriy H Nevidomskyy and Hitesh J Changlani 2022 *npj Quantum Materials* **7** 51.

24) Hirano Y, Skantha kumar S, Loong C K, Wakabayashi N, and Boatner L A 2002 *Phys. Rev. B* **66** 024424.

25) Suzuki H, Higashino Y and Inoue T 1980 *J Phys. Soc. Vol.* **49**, 3.

26) Kirschbaum K, Martin A, Parrish D A and Pinkerton A A 1999 *J Phys. Condens. Matter* **11** 4483.

27) Becker P J, Dummer G, Kahle H G, Klein L, Muller-Vogt G, H C 1970 *Phys. Letts.* A **31** 499.

28) Will G and Schafer W 1971 *J. Phys. C: Solid State Phys.* **4** 811.

29) Massat P, Wen J, Jiang J M, Hristov A T, Liu Y, Feigelson R S, Lee Y S, Fernandes R M and Fisher I R 2022 *PNAS vol.* 119, No. **28**, 2119942119.

30) Vinograd I, Shirer K R, Massat P, Wang Z, Kissikov T, Garcia D, Bachmann M D, Horvatic M, Fisher I R and Curro N J 2022 *npj Quantum Mater.* **7,** 68.

31) Wang Z, Vinograd I, Mei Z, Menegasso P, Garcia D, Massat P, Fisher I R, and Curro N J 2021 *Phys Rev B* **104** 205137.

32) Ranaut D and Mukherjee K 2022 *Sci. Rep.* **12** 56.

33) Ranaut D and Mukherjee K 2022 *J. Phys.: Condens. Matter* **34** 315802.

34) Battison J E, Kastent A, Leask M J M and Lowry J B 1977 *J. Phys. C: Solid State Phys.* **10** 323.

35) Kondrateva O N, Nikiforova G E, Tyurin A V, Smirnova M N, Gavrichev K S 2020 *J. Alloys and Compounds* https://doi.org/10.1016/j.jallcom.2020.157786.

36) Skanthakumar S, Loong C K and Soderholm L 1995 *Phys Rev B* **51** 18.

37) Nicklow R M, Moon R M, Kawarazaki S, Kunitomi N, Suzuki H, Ohtsuka T and Morii Y 1985 *J. Appl. Phys.* **57** 3784.

38) Bleaney B, Bowden G J and Clark R G 1982 *J. Phys. C: Solid State Phys.* 1982 **15** 1663.

39) http://elk.sourceforge.net;

40) Lal S and Pandey S K 2017 *Phys. Lett. A* **381** 917.

41) Perdew J P, and Wang Y 1992 *Phys. Rev. B.* **45** 13244.

42) Pakhira S, Mazumdar C, Ranganathan R and Avdeev M 2017 *Sci. Rep.* **7** 7367.

43) Itou T, Oyamada A, Maegawa S, Tamura M, and Kato R 2008 *Phys. Rev. B* **77** 104413.

44) Shimizu Y, Miyagawa K, Kanoda K, Maesato M, and Saito G 2003 *Phys. Rev. Letter* **91** 107001.

45) Okamoto Y, Nohara M, Aruga-Katori H, and Takagi H 2007 *Phys. Rev. Letter* **99** 137207.

46) Choi Y. S., Lee C. H., Lee S., Yoon S, Lee W J, Park J, Ali A, Singh Y, Orain J C, Kim G, Rhyee J, Chen W T, Chou F and Choi K Y 2019 *Phys. Rev. Letts.* **122** 167202.

47) Mykola Abramchuk, Cigdem Ozsoy-Keskinbora, Jason W. Krizan, Kenneth R. Metz, David C. Bell and Fazel Tafti 2017 *J. Am. Chem. Soc.* **139** 15371.

48) Kittel C, Introduction to Solid State Physics (Wiley, Singapore, 1996).

49) Calder S, Ke X, Bert F, Amato A, Baines C, Carboni C, Cava R J, Daoud-Aladine A, Deen P, Fennell T, Hillier A D, Karunadasa H, Taylor J W, Mendels P, Schiffer P and Bramwell S T 2010 *Phys. Rev. B* **81** 064425.

50) Mohapatra N, Iyer K. K., Rayaprol S., and Sampathkumaran E.V. 2007 *Phys. Rev. B* **75**, 214422.

51) Amor N B, Bejar M, Dhahri E, Valente M A, Lachkar P, Hill E K 2013 *J. Rare Earths* **31** 54.

52) Lounasmaa O. V. 1962 *Phys. Rev.* **128** 1136.

53) Cheng J. G, Li G, Balicas L, Zhou J. S., Goodenough J. B., Xu C., and Zhou H. D. 2011 *Phys. Rev. Letts.* **107** 197204.

54) Moussa M, Djermouni M, Kacimi S, Azzouz M, Dahani A, Zaoui A 2013 *Comput. Mater. Sci.* **68** 361.

55) E.S.R. Gopal. Specific Heat at Low Temperatures. Department at Physics, Indian Institute at Science Bangalore, India.

**Figures:**

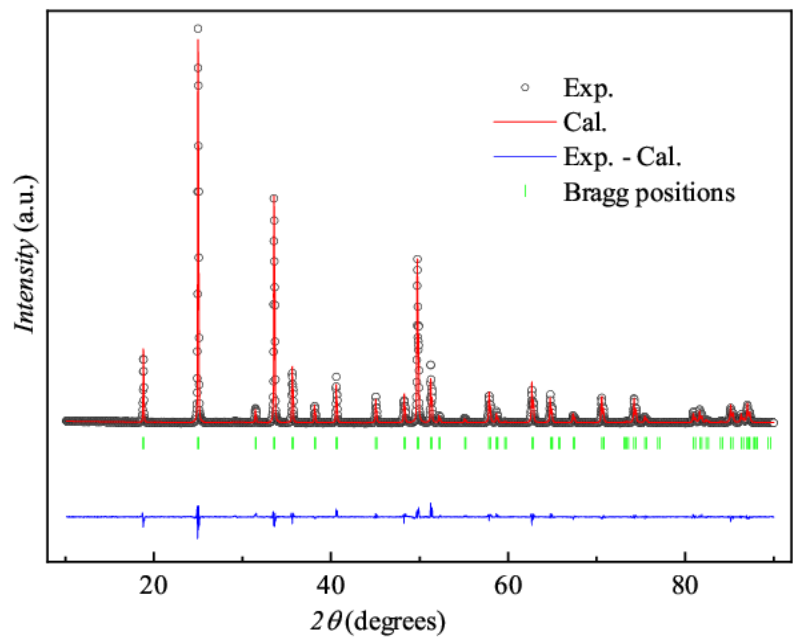

**Fig. 1.** Rietveld refined powder x-ray diffraction pattern of HoVO$_4$. The black open circles indicate to the observed data, while the Rietveld refinement fit is shown as red solid line. The difference curve and the Bragg positions are shown by blue solid line and green vertical lines, respectively.

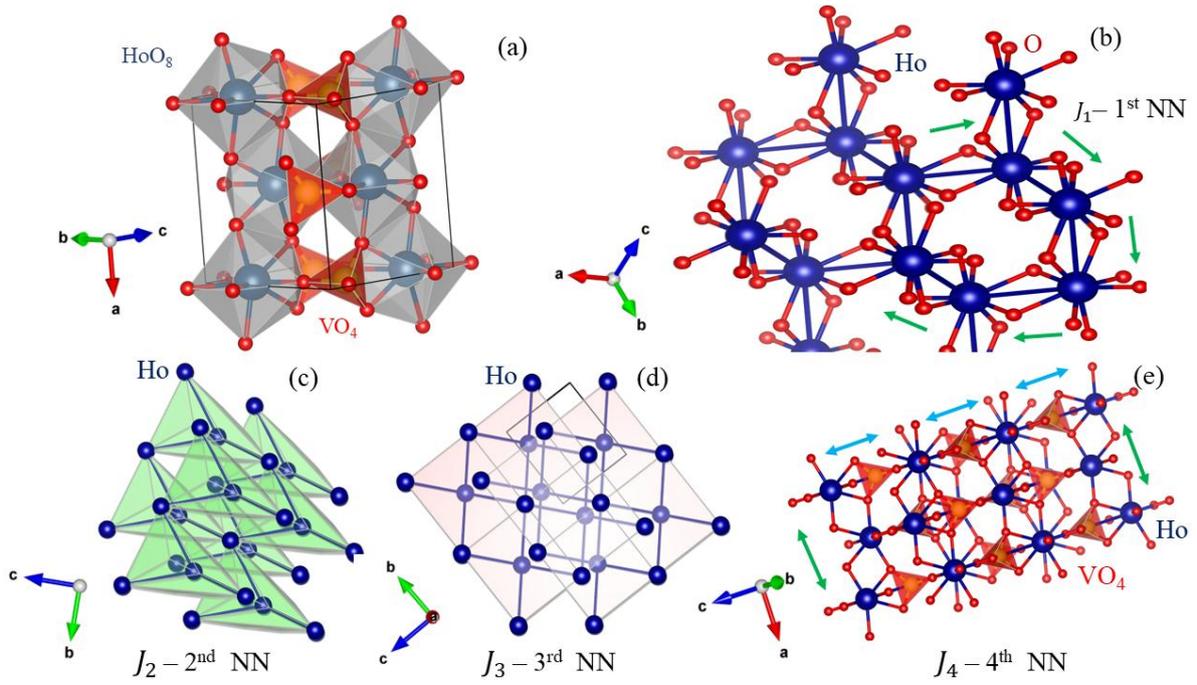

**Fig. 2.** (a) The crystal structure of $HoVO_4$ comprised of $HoO_8$ polyhedral (grey color) and $VO_4$ tetrahedra (orange color). (b) The first NNs Ho atoms form a 3D network of spins. The green arrow shows the direction of interaction path. (c) Formation of a frustrated network of corner-sharing tetrahedra by second NNs Ho atoms. (d) The third NNs Ho atoms form a frustrated network of edge-sharing polyhedral. (e) The fourth NNs Ho atoms make uniform chains, shown by blue arrow. These chains are interlinked via $J_1$, shown by the green arrow.

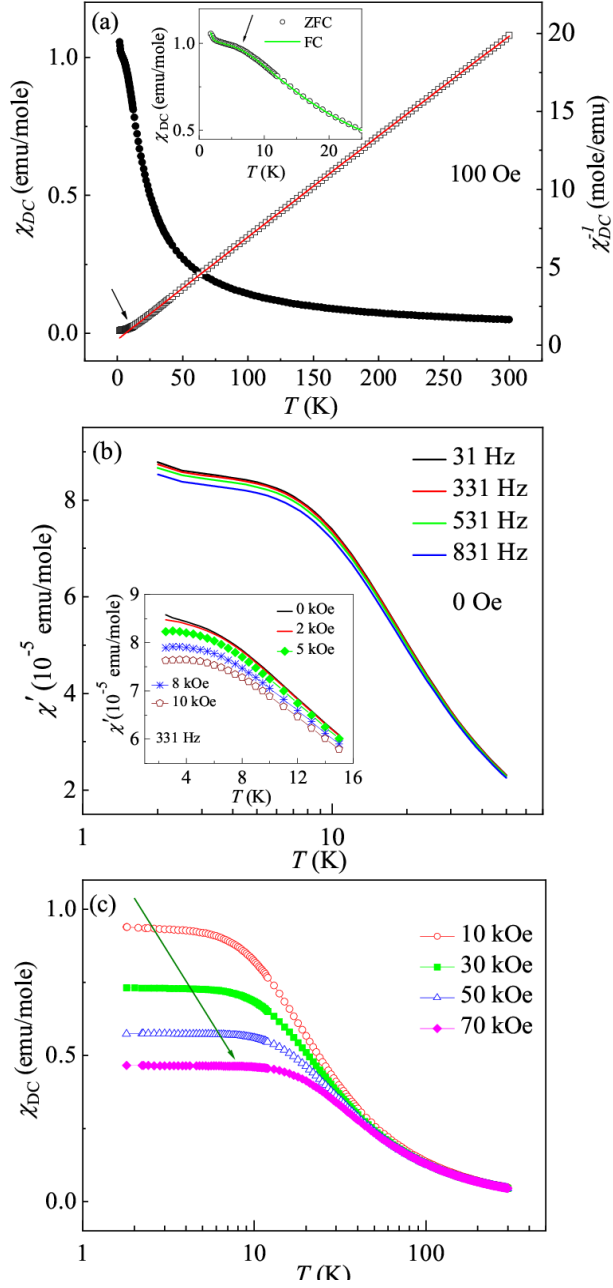

**Fig. 3.** (a) $\chi_{DC}$ (T) vs. T under ZFC condition (left panel) and $\chi_{DC}^{-1}$ vs. T at 100 Oe (right panel). The red solid line shows the CW fitting with arrow showing the deviation from the fitted line. The inset shows the expanded low temperature for both ZFC and FC curves at same field. The arrow in the inset shows the position of broad hump. (b) T dependence of $\chi'$ at various frequencies for 0 Oe. The inset shows $\chi'(T)$ at various applied DC fields corresponding to AC frequency 331 Hz. (c) $\chi_{DC}$ (T) vs. T at different applied H up to 70 kOe on a log scale. The green arrow shows the onset of temperature independent behaviour at different H.

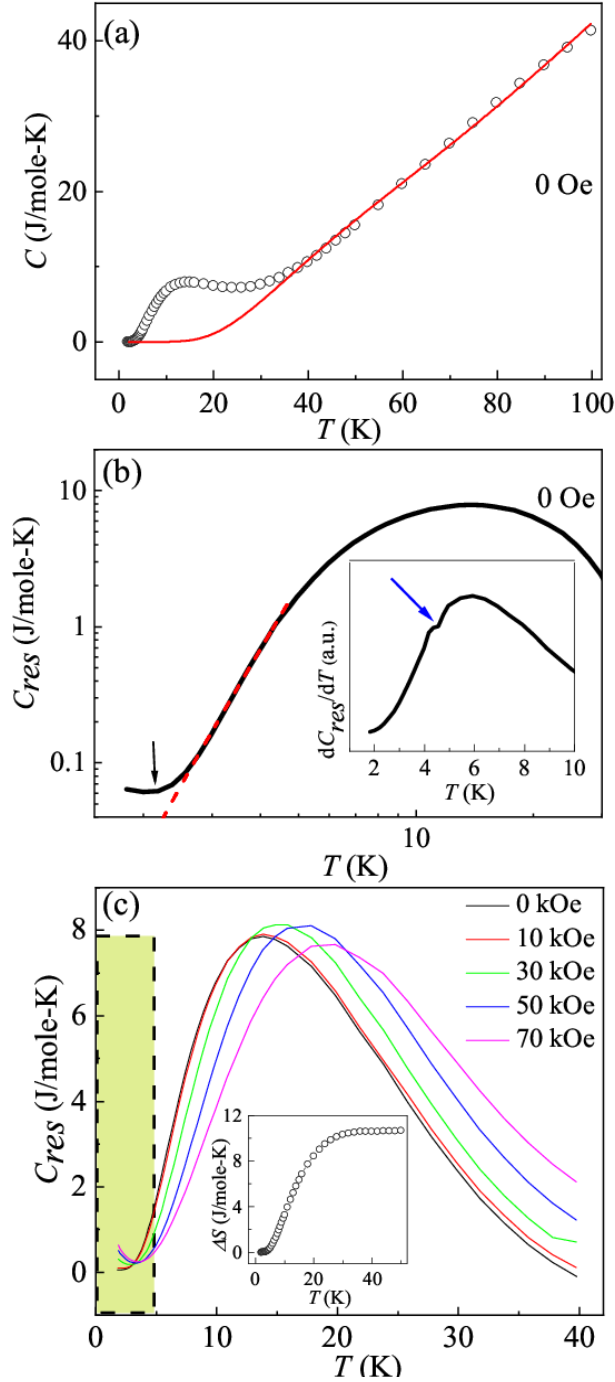

**Fig. 4.** (a) $C$ vs $T$ in zero field with the fit to the equation (2), shown by the red solid line. (b) The $T$ dependent $C_{res}$ curve at zero field on a log-log scale. The red dashed line shows the fit to power-law ($C_{res} \propto T^{\alpha}$) with $\alpha \sim 2.2$. The black solid arrow shows the evolution of nuclear Schottky anomaly. The inset shows the $dC_{res}/dT$ vs $T$ curve at zero field and the arrow shows the presence of an anomaly. (c) $C_{res}$ as a function of $T$ measured under different external $H$. The yellow shaded region shows the development of nuclear Schottky anomaly under the application of $H$. The inset shows the entropy change ($\Delta S$) calculated from integrating $C_{res}/T$ vs $T$ curve above 1.8 K at zero applied magnetic field.

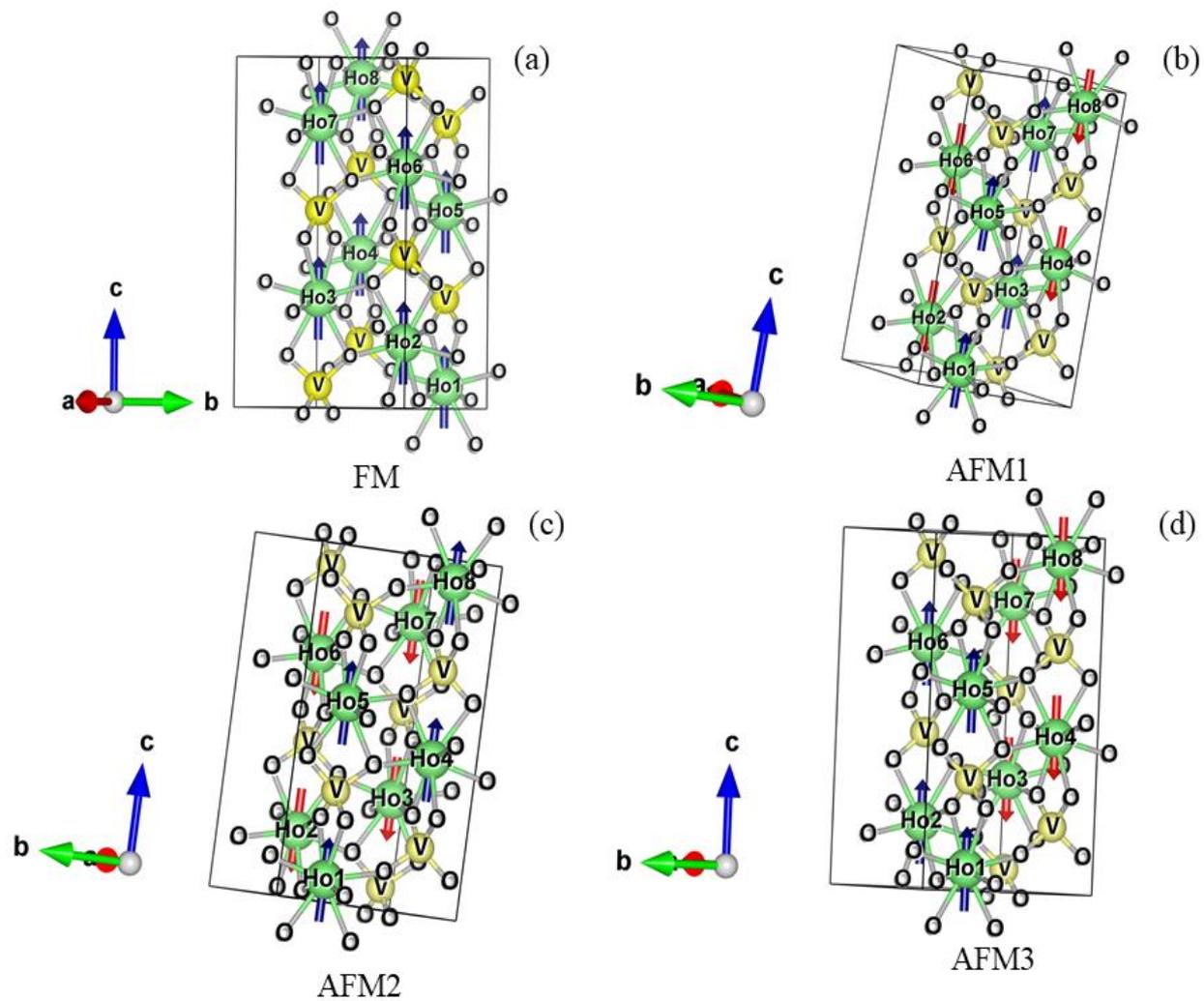

**Fig. 5** The magnetic structures corresponding to different spin configurations. (a) FM spin configuration. (b-d) Different AFM spin configurations.

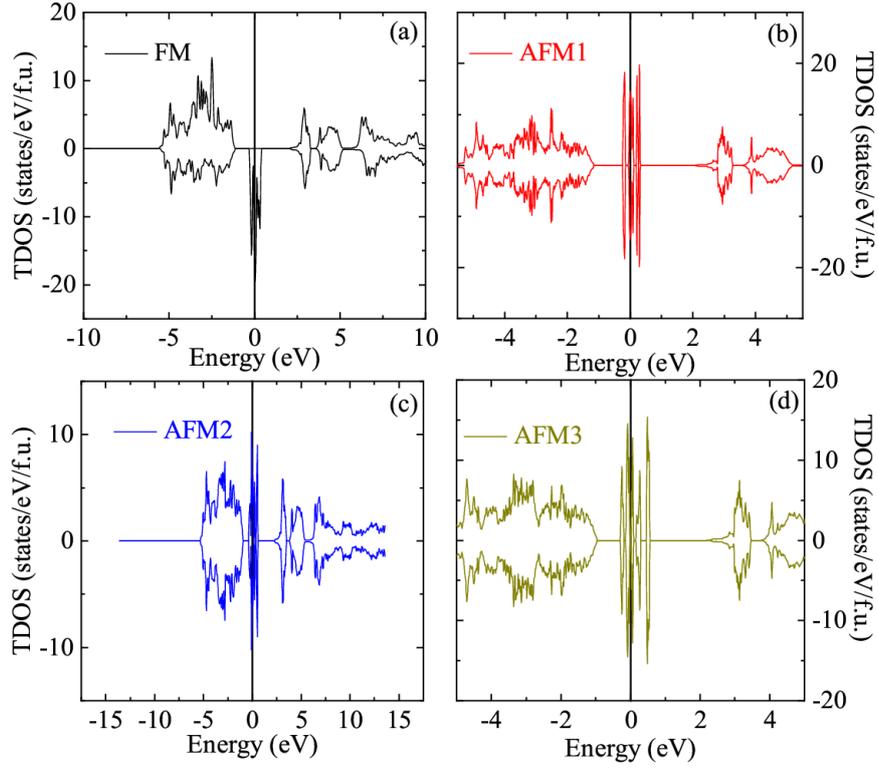

**Fig. A1:** TDOS for different structures obtained under the DFT formalism. The black lines passing through origin denotes the Fermi energy.

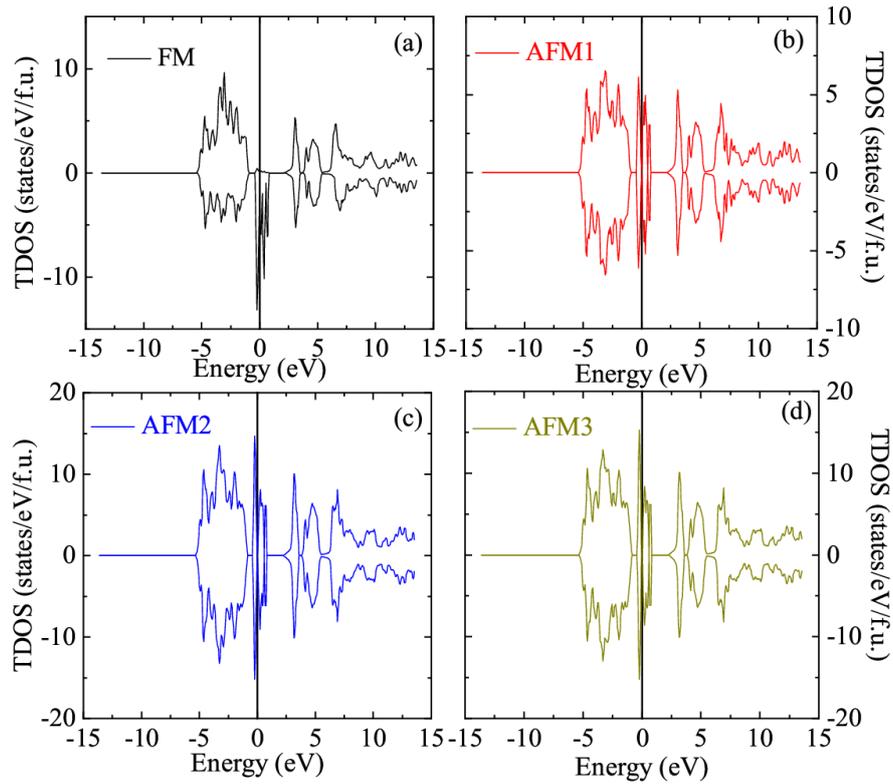

**Fig. A2:** TDOS for four different magnetic structures obtained under DFT+SOC formalism. The black lines passing through origin denotes the Fermi energy.